\documentclass[copyright]{eptcs}
\providecommand{\event}{T. Neary, D. Woods, A.K. Seda and N. Murphy (Eds.):\\ The Complexity of Simple Programs 2008.}
\providecommand{\volume}{1}
\providecommand{\anno}{2009}
\providecommand{\firstpage}{22}
\usepackage{breakurl}

\usepackage{amsmath}
\usepackage[latin1]{inputenc}
\usepackage{amstext,amsopn,eucal}
\usepackage{graphics}
\usepackage{latexsym}
\usepackage{exscale}
\usepackage{graphicx}
\usepackage{amssymb} 
\usepackage{amsfonts}
\usepackage{graphics}
\usepackage{xspace}

\newtheorem{theorem}{Theorem}[section]

\newtheorem{example}{Example}[section]

\newtheorem{corollary}{Corollary}[section]

\newcommand{\RE}{\ensuremath{\sf RE}}
\newcommand{\CF}{\ensuremath{\sf CF}}
\newcommand{\REG}{\ensuremath{\sf REG}}

\newcommand{\lra}{\longrightarrow}
\newcommand{\ra}{\rightarrow}
\newcommand{\Ra}{\Rightarrow}

\newcommand\upr[2]{\left(\!\begin{tabular}{c}$#1$\\$#2$\end{tabular}\!\right)}
\newcommand\ups[2]{\left[\!\begin{tabular}{c}$#1$\\$#2$\end{tabular}\!\right]}

\newcommand\hiover{
\put(10,10){\line(1,0){60}}
\put(40,20){\line(1,0){30}}
\put(10,30){\line(1,0){30}}
\put(10,10){\line(0,1){20}}
\put(40,20){\line(0,1){10}}
\put(70,10){\line(0,1){10}}
\put(20,15){u}}
\newcommand\loover{
\put(10,10){\line(1,0){30}}
\put(40,20){\line(1,0){30}}
\put(10,30){\line(1,0){60}}
\put(10,10){\line(0,1){20}}
\put(40,10){\line(0,1){10}}
\put(70,20){\line(0,1){10}}
\put(20,15){u}}
\newcommand\strand[1]{
\put(10,10){\framebox(#1,10){v}}} 

\title{Complexity through the Observation of Simple Systems}

\author{Matteo Cavaliere
    \institute{CoSBi\\ Trento, Italy}
    \email{cavaliere@cosbi.eu}
    \and
    Peter Leupold
    \institute{Department of Mathematics, Faculty of Science\\
 Kyoto Sangyo University  \\ Kyoto 603-8555, Japan\thanks{This work was done while Peter Leupold was funded as a post-doctoral fellow by the Japanese Society for the Promotion of Science under number P07810.}}
    \email{leupold@cc.kyoto-su.ac.jp}
}

\def\titlerunning{Complexity through the Observation of Simple Systems}
\def\authorrunning{M. Cavaliere and P. Leupold}

\begin{document}
\maketitle

\begin{abstract} 
We survey work on the paradigm called ``computing by observing." Its central feature is that one considers the behavior of an evolving system as the result of a computation. To this end an observer records this behavior. It has turned out that {\em the observed behavior of computationally simple systems can be very complex}, when an appropriate observer is used. For example, a restricted version of context-free grammars with regular observers suffices to obtain computational completeness. As a second instantiation presented here, we apply an observer to sticker systems. Finally, some directions for further research are proposed.  
\end{abstract}

\section{Introduction}

In the standard input-output paradigm in Computer Science, the result of a computation is obtained by introducing an input to some type of system, waiting while the system processes this input, and then reading the produced result. For example, a Turing Machine changes the contents of its tape according to its program until it reaches a final state; then a specified portion of the tape contains the output. The process that the system runs through and by which it produces the result is in general completely ignored. 

``Computing by observing" uses a different approach: the entire process of the system is actually the produced result. The final or initial state by themselves are irrelevant. This approach supposes the presence of an external observer that is able to watch the system during its functioning and to extract a formal recording of its behavior. Thus in this paradigm, a computing device is constructed from two main components: the first one called the basic or underlying system and a second one, called the observer.  The paradigm is inspired from the procedure followed in experimental sciences. There, it is a standard proceeding to conduct an experiment, observe the entire progress of a system, and then take the result of this observation as the final output. Consider for example the use of a catalyst in a chemical reaction. The reaction might produce the same result, but the energy used or the time required might be less. This difference only becomes clear when looking at a protocol of how the quantities in question evolved during the course of the experiment.  

\begin{figure}[h]
     \begin{center}
     \includegraphics[scale=1.0]{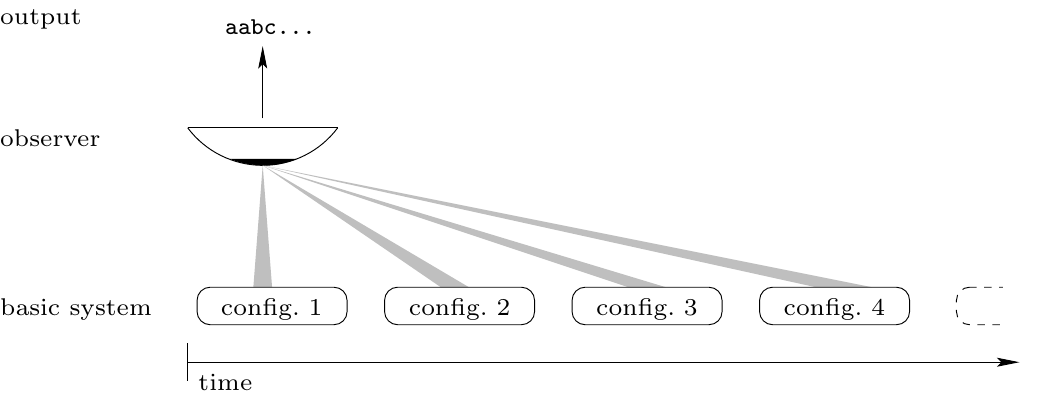}
     \end{center}
    \caption{Basic structure of the ``Computing by Observing" architecture.}
\label{idea}
\end{figure}

The architecture that protocols the evolution of a system in the way described is schematically depicted in Figure \ref{idea}.

The approach stresses the role of the observer in computation. This role can be crucial as we will see in Section \ref{COG}: in fact, any computational device can be obtained by observing in the right manner a fixed (and simple) basic system (one can then talk of ``computing by only observing...").

In what follows we will use some basic notions and standard terminology from formal languages theory (see, e.g., \cite{Ullman}). 
We only mention that  $\mathcal{REG}$ and $\mathcal{CF}$ denote the classes of regular and context-free grammars, respectively and that by $\REG$, $\CF$ and $\RE$ we denote the classes of languages generated by regular, context-free, and type-0 grammars, respectively.

\section{Observed Complexity of Simple Grammars} \label{COG}

In this section we present {\em grammar/observer (G/O) systems} that are generative devices based on the  ``computing by observing" paradigm.
In this case, a formal grammar plays the role of the observed basic system and the external observer is a finite automaton.
This section is essentially based on \cite{CL05}.

\subsection{The Observers: Monadic Transducers} \label{ofl}

A grammar's states are the sentential forms of its derivations. So for the observer we need a device mapping these arbitrarily long strings into just one singular symbol. We use a special variant of finite automata with some feature known from Moore machines or also from subsequential transducers: the set of states is labeled with the symbols of an output alphabet. Any
computation of the automaton produces as output the label of the state it halts in. Because we find it preferable that the observation of a certain string always leads to a fixed result, we consider here only deterministic and complete automata. The name monadic transducer is motivated from monadic string-rewriting rules; these can have arbitrarily long left sides that are rewritten to strings of length one or zero.

For simplicity, in what follows, we present only the mappings that the observers define, without giving a real implementation (in terms of finite automata) for them.  Therefore no more formal definition of monadic transducers is necessary here. The class of all monadic transducers is denoted by $FA_O$.

\subsection{Grammar/Observer Systems} \label{ofg}

A {\em Grammar/Observer (G/O) system} is a pair $\Omega = (G,A)$ constituted by a generative grammar $G =(N, T,  S, P)$ and a monadic transducer (observer) $A$ with output alphabet $\Sigma \cup \{\bot\}$, which then is also the output alphabet of the entire system $\Omega$. The transducer's input alphabet must be the union of $N$ and $T$ from the grammar so that it can read all sentential forms.

Several modes of generation can be defined (see, e.g., \cite{CL05}). Here we will consider the mode of generation that admits writing an empty and non-empty output in an arbitrary manner ({\em free G/O systems}); i.e., the transducer can either output one letter or the empty word and freely alternate between these two options.
A {\em free G/O system} generates a language in the following manner:  
 \begin{eqnarray*} L_f(\Omega) &=& \{A(w_0,w_1,\dots ,w_n) \mid S=w_0 \Rightarrow w_1 \Rightarrow \dots \Rightarrow w_n,\  w_n\in T^*  \}.\end{eqnarray*}
Here $A(w_0,w_1,\dots ,w_n)$ is used as a more compact way of writing the catenation $A(w_0)A(w_1)\cdots A(w_n)$. 

In other words, the language contains all those words which the observer can produce during the possible terminating derivations of the underlying grammar. Derivations which do not terminate do not produce a valid output; this means that we only take into account finite words. Of course, by considering the other case of non-terminating derivations the G/O systems could also be used to generate languages of infinite words.

We also consider the variant where we define the language produced by $\Omega$ as 
$$ L_{\bot,f}(\Omega) =  {L}_f(\Omega) \cap \Sigma^* .$$
In this way the strings in $L_f(\Omega)$ containing $\bot$ are filtered out and they are not present in $L_{\bot,f}(\Omega)$. Thus the observer has in some sense the ability to reject a computation, when configurations of a certain class appear.
The language generated by $\Omega$ is also called observed behavior of $\Omega$.

The functioning of a (free) G/O system is sketched in Example \ref{EXderivat}.

\begin{example} \label{EXderivat}

\begin{figure}[ht]\begin{center}
    \scalebox{0.67}{\includegraphics{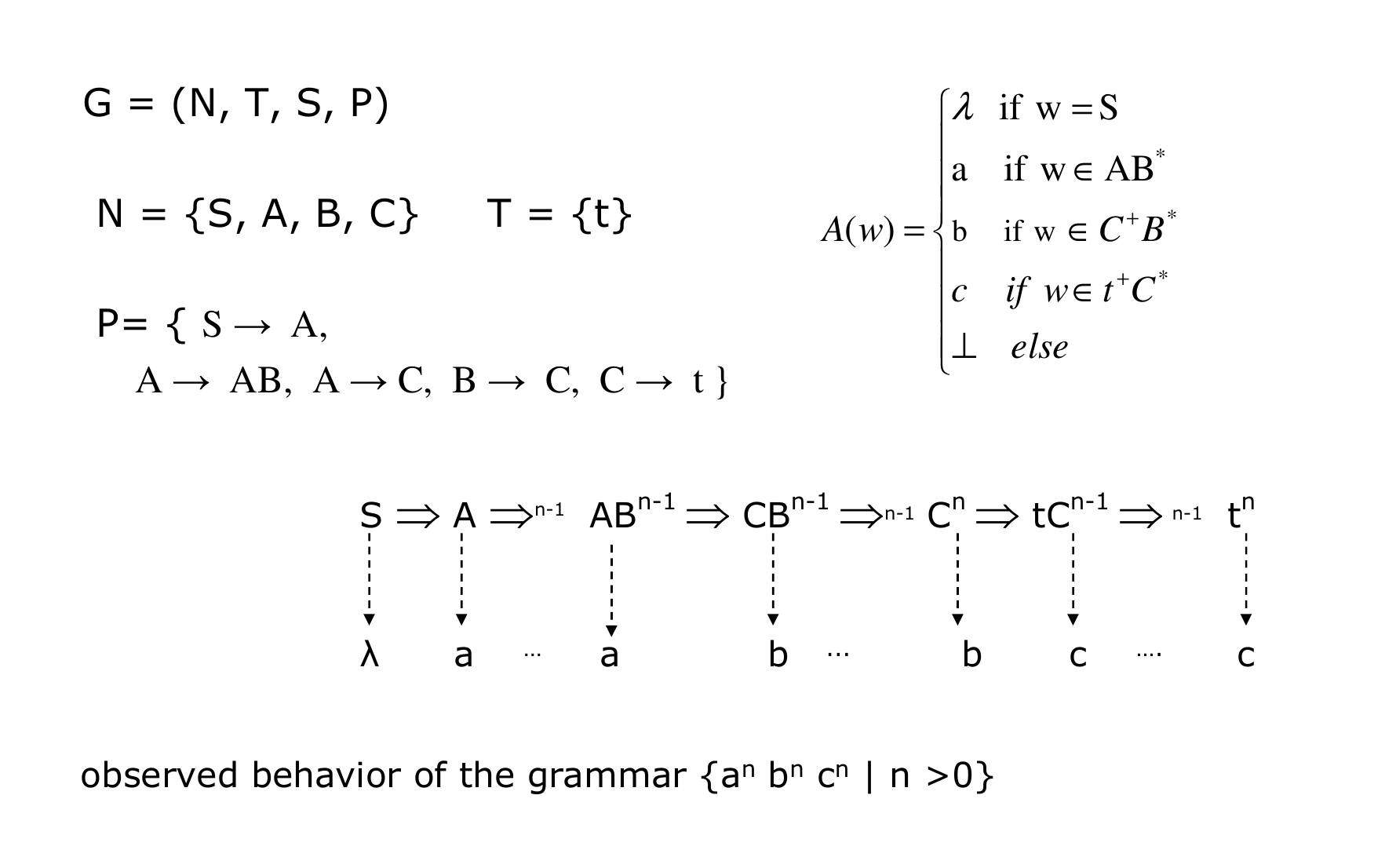}}\end{center}
 \end{figure}
 \end{example}
To each sentential form produced by the grammar $G$ the observer $A$ associates a symbol that can be $a,b,c,\bot$ or the empty string $\lambda$ (the vertical arrow is the observer mapping). Thus every computation of the observer produces one output symbol and the concatenation of these symbols is then the output string. For instance, in the figure the output string is $\lambda a \cdots a b\cdots b c \cdots c$. The mapping defined by the observer is specified by the regular expressions. The language $L_f(\Omega)$ is obtained by considering all possible halting derivations of $G$ and collecting all the output strings. It is easy to see that in this case $L_{\bot,f} (\Omega)$ is $\{a^n b^n c^n \mid n >0\}$.

In \cite{CL05} it has been shown that a $G/O$ system composed of very simple  components, namely a locally commutative context-free grammar ($LCCF$), a proper subclass of context-free grammars, and a finite state automaton is computationally complete.

\begin{theorem}\label{tRE2} For each $L \in RE$ there exists a G/O system $\Omega = (G,A)$, with $G$ an $LCCF$, such that $L_{\bot,f}(\Omega)=L.$  \cite{CL05} \end{theorem}
 
An interesting fact is that the observer's ability to produce $\bot$, i.e., to eliminate certain computations, seems to be a powerful and essential feature in all the variants explored in \cite{CL05}. Besides the free variant presented here, also systems that have to write in every single step and systems that have to write in every step after they have started to write are investigated there. 

As can be seen from the definition, for free systems we obtain all recursively enumerable languages over $\Sigma$ simply by intersection of a language over $\Sigma \cup \{\bot\}$ with the regular language $\Sigma^*$.  Now notice that recursive languages are closed under intersection with regular sets. Therefore, there must exist some  grammar/observer systems $\Omega$  generating a {\em non-recursive} $L_{f}(\Omega)$ (i.e., not using the filtering with $\bot$). However, this intersection filters out a great many words produced from undesired computations; so despite being simple, this intersection distinguishes between good and bad computations. Therefore it seems very unlikely that the same model without this feature will be computationally complete, although it generates non-recursive languages.

As discussed in the introduction, the idea of the presented framework is to stress the role of the observer in computations. Therefore, it is interesting to understand how much one can compute by making only changes in the observer, keeping  the observed basic system unchanged.  We show by means of an example that a G/O system can generate very different languages if the observer is changed while the grammar remains fixed.

Let us consider the context-free grammar $G = (\{S,A,B,C\}$,$\{t,p\},S$,$\{S \ra pS, S \ra p, S\ra A, A\ra AB, A\ra C, B\ra C, C\ra t\})$.
If $G$ is coupled with the observer $A'$ such that\linebreak 
$A'(w) = a \textrm{ if } w\in \{S,A,B,C,t,p\}^+ $,
then $\Omega=(G,A')$ defines the language $L_{\bot,f}(\Omega)= \{a^i \mid i \geq 2\}$, a regular language. 
In fact, the derivation $S\ra pS \stackrel{n-2}\Ra p^{n-1}S \ra p^n$ produces (when observed) the string $a^{n+1}$.

\smallskip
Keeping the same grammar $G$ we change the observer into $A''$ such that:
$$A''(w) = \left\{
 \begin{array}{ll} 
    \lambda & \textrm{ if } w=S,\\
    a & \textrm{ if } w\in AB^*,\\
    b & \textrm{ if } w\in C^+B^*,\\
    c & \textrm{ if } w\in t^+C^*,\\
    \bot & \textrm{ else}
 \end{array}\right.$$
In this case, one can verifies that $\Omega=(G,A'')$ generates the language $L_{\bot,f}(\Omega)= \{a^nb^nc^n \mid n > 0 \}$, a context-sensitive language. 

This example suffices to underline that part of the computation can be done by choosing the right observer, keeping unchanged the underlying basic system.
 Actually, this part can be really crucial: As shown in \cite{COO} one can construct an {\em universal context-free grammar} that can generate all recursively enumerable languages when observed in the correct manner.

We can also consider several {\em restrictions on the observed system}. In particular, we can bound the number of nonterminals in the sentential forms. 
 
In this respect, we notice that the universal context-free grammar used in \cite{COO} has no bound on the number of nonterminals
in its sentential forms.

The next result shows that indeed this is
a necessary property of context-free grammars that are
observationally complete for type-0 grammars.
In fact, when a bound is imposed, the observed behaviors  are regular. 
Recall that a context-free grammar is {\em nonterminal bounded}
if there exists a constant $k$ such that all sentential forms
generated by the grammar have at most $k$ nonterminals. Clearly, a regular grammar is nonterminal bounded.

\begin{theorem}
\label{th-ultra}
For every G/O system $\Omega = (G,A)$, with $G$ nonterminal bounded context-free,  
$L_{\bot,f}(\Omega)$ is regular. \cite{COO}
\end{theorem}

\section{Observed Complexity of Simple BioSystems}

 As we have seen in Section \ref{ofg} the complexity of the produced output is influenced by the particular dynamics of the observed system. This means that the entire trajectory followed by the observed system is important rather than the momentary reached states.  In this section we stress the distinction between the complexity of the result and of the behavior by considering as observed system a sticker system  (a computational model inspired by the self-assembly of DNA strands).  It is known that  sticker systems have a computational power that is less than that of regular grammars. However their behavioral complexity is computationally complete when observed in an appropriate manner. This means that their observed behaviors can represent all possible computable languages. In other words, a basic system (sticker system) that is less powerful than a regular grammar, is then universal in terms of observed behaviors ({\em this result clearly contrasts with Theorem \ref{th-ultra}: observed behaviors of regular grammars are still regular}).

 Sticker systems were introduced in \cite{stickefirst} as a formal model of the operation
of {\em annealing} (and {\em ligation}) operation that is  largely used in 
DNA computing area.  The basic operation of a sticker
system is the {\em sticking} operation that constructs double
stranded sequences out of ``DNA dominoes" ({\em polyominoes}) that
are sequences with one or two sticky ends, or
single stranded sequences, attaching to each other by {\em
ligation} and {\em annealing}.

An {\em observable sticker system} was  then introduced in \cite{AC04}. The idea of an observable sticker system can be
expressed in the following way: an observer (for example, a
microscope) is placed outside the ``test tube", where (an
unbounded number of copies of) DNA strands and DNA dominoes are placed
together. Some of these molecules are marked (for example, with a
fluorescent particle). The molecules in the solution will start to
self-assemble (to stick to each other) and, in this way, new
molecules are obtained. The observer watches the trajectory of
the marked molecules and stores such evolution on an external tape
in a chronological order.
For each possible trajectory of the marked molecules a certain
string is obtained. Collecting all the possible trajectories of
such marked strands we obtain a language.

We first recall the definition of sticker systems and we couple them with an external observer, defining  an observable sticker system. This part is essentially based on \cite{AC04}.

Consider a symmetric relation
$\rho \subseteq V \times V$ over $V$ (of {\em complementarity}).
Following \cite{DnaBook}, we associate with $V$
the monoid $V^* \times V^*$ of pairs of strings.
Because it is intended to represent DNA molecules, we also write elements
$(x_1,x_2) \in V^* \times V^*$ in the form $\upr{x_1}{x_2}$ and $V^* \times V^*$ as $\upr{V^*}{V^*}$.
We denote by $\ups{V}{V}_\rho=\{\ups{a}{b}\mid a,b\in V, (a,b)\in\rho\}$
the set of {\em complete double symbols},
and $WK_\rho(V)=\ups{V}{V}_\rho^*$ is the set of the
{\em complete double-stranded sequences (complete molecules)}
also written as $\ups{x_1}{x_2}$, where $x_1$ is the {\em upper strand}
and $x_2$ is the {\em lower strand}.

\begin{figure}[hb]\begin{center}
\begin{picture}(340,30)(0,30)
\put(0,0){\hiover}\put(62,0){\strand{40}}
\put(0,30){\loover}\put(62,40){\strand{40}}
\put(130,0){\hiover}\put(162,12){\strand{20}}
\put(130,30){\loover}\put(162,28){\strand{20}}
\put(230,0){\hiover}\put(262,12){\strand{60}}
\put(230,30){\loover}\put(262,28){\strand{60}}
\end{picture}\end{center}
\vspace{10mm}
\caption{Sticking operation\label{Pstick}}
\end{figure}
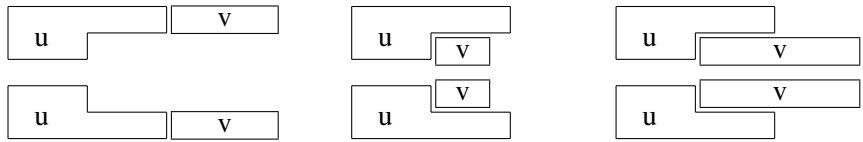

As in \cite{DnaBook}, we use {\em single strands} -- the
elements of $S(V)=\upr{\lambda}{V^*}\cup\upr{V^*}{\lambda}$ and
the molecules with (a possible) {\em overhang} on the right, which
are the elements of $R_\rho(V)=\ups{V}{V}_\rho^*S(V)$, from now on
called {\em well-started} molecules (upper and lower strand are
defined as in the case of complete molecules).

Given a well started molecule $u\in R_\rho(V)$ and a single strand
$v\in S(V)$, we recall in Figure \ref{Pstick} the partial operation
$\mu:R_\rho(V)\times S(V)\lra R_\rho(V)$ of sticking, as defined in \cite{DnaBook}.
We point out that we use a case of sticking, restricted to pasting a single strand
to the right side of a well-started molecule (with a possible overhang on the right),
corresponding to the {\em simple regular} sticker systems.
  
A (simple regular) sticker system is a construct
$\gamma=(V,\rho,X,D)$, where
$X\subseteq R_\rho(V)$ is the (finite) set of axioms,
and $D\subseteq S(V)$ is the (finite) set of {\em dominoes}
(in this case these are single strands).
Given $u,w\in R_\rho(V)$, we write $u\Ra w$ iff $w=\mu(u,v)$ for some $v\in D$.
A sequence $(w_i)_{1\leq i\leq k}\subseteq R_\rho(V)$
is called a {\em complete computation} if $w_1\in X$, $w_i\Ra w_{i+1}$ for $1\leq i< k$ and $w_k\in WK_\rho(V)$.

The {\em language} generated by a sticker system $\gamma$ is the
set of upper strands of all complete molecules derived from the
axioms.
It is known that the {\em family of languages generated by
simple regular sticker systems is strictly included in the family
of regular languages} (see \cite{DnaBook} for the proof).

For an alphabet $V$, our {\em double-symbol alphabet} constructed over $V$ is
$$V_d=\ups{V}{V}_\rho\cup \upr{V}{\lambda}\cup\upr{\lambda}{V}.$$
We define an observer $A \in FA_O$, with input alphabet $V_d$, that reads an entire molecule (element of $R_\rho(V)$)
and outputs one symbol from the output alphabet $\Sigma \cup \{\bot\}$ (every well-started molecule in $R_\rho(V)\subseteq V_d^*$ is read,
in a classical way, from left to right, scanning one double symbol from $V_d$ at a time).
For a molecule $w\in R_\rho(V)$ and the observer $A$ we write $A(w)$ to indicate such output;
for a sequence $w_1,\dots, w_n$ of $n\geq1$ of molecules in $R_\rho(V)$
we write $A(w_1,\dots ,w_n)$ for the string $A(w_1)\cdots A(w_n)$.

An {\em observable sticker system} with output alphabet $\Sigma \cup \{\bot\}$
is a construct $\phi=(\gamma,A)$, where $\gamma$ is the sticker
system with alphabet $V$, and $A \in FA_O$ is the observer with input
alphabet $V_d$ and  output alphabet $\Sigma \cup \{\bot\}$.

We denote the collection of all complete computations of $\phi$ by $\mathcal{C}(\phi)$.
The language, over the output alphabet $\Sigma \cup \{\bot\}$, generated by an observable sticker system $\phi$,
is defined as $L(\phi)= \{A(s)\mid s \in \mathcal{C}(\phi)\}$.
As done for G/O systems, we want to filter out the words that contain the special symbol $\bot$, then we consider
the language $L_\bot(\phi)= {L}(\phi) \cap \Sigma^*$.

We will illustrate with a simple example how an observable sticker
system works. At the same time this example shows how one can
construct an observable sticker system with an observed behavior that is a non regular
language despite the fact that the power of simple regular
sticker systems, when considered in the classical way, is
subregular.

Consider the following observable sticker system $\phi=(\gamma,A)$:
\begin{eqnarray*}
\gamma&=&(V=\{a,{\bf c},{\bf g},t\},\rho=\{(a,t),({\bf c},{\bf g}),(t,a),({\bf g},{\bf c})\},
X=\{\ups{a}{t}\},D),\\
D&=&\{\upr{a}{\lambda},\upr{\lambda}{t},\upr{\bf c}{\lambda},\upr{\lambda}{\bf g}\},
\end{eqnarray*}
with the observer $A$ defined by the following mapping:
\begin{center}
$A(w)=\left\{
 \begin{array}{llll}
b,&\textrm{ if }&w\in&\ups{a}{t}^*\upr{a^*}{\lambda}\cup\upr{\lambda}{t^*},\\
d,&\textrm{ if }&w\in&\ups{a}{t}^*\upr{a^*{\bf c}}{\lambda}\cup\upr{\lambda}{t^*{\bf g}},\\
\lambda,& &&\textrm{otherwise}.
\end{array}\right.$
\end{center}
The language generated by $\gamma$ is $L_1=\{b^md^n\mid m\geq n, m\geq 1, n\geq 0\}\notin REG$.

Below is an example for a computation of $\phi$ (generating $bbbbdd$):

\begin{center}%
\noindent\begin{tabular}{|l|l|l|l|l|l|l|l|l|}\hline
Step&0&1&2&3&4&5&6\\ \hline
Added&&$\upr{a}{\lambda}$&$\upr{a}{\lambda}$&$\upr{\lambda}{t}$
&$\upr{\bf c}{\lambda}$&$\upr{\lambda}{t}$&$\upr{\lambda}{\bf g}$\\ \hline
Molecule&$a$&$aa$&$aaa$&$aaa$&$aaa\bf c$&$aaa\bf c$&$aaa\bf c$\\
&$t$&$t$&$t$&$t\,t$&$t\,t$&$t\,t\,t$&$t\,t\,t\,\bf g$\\ \hline
Output&$b$&$b$&$b$&$b$&$d$&$d$&$\lambda$\\ \hline
\end{tabular}
\end{center}%

The idea of the system $\phi$ is the following: think of symbols {\bf c}, {\bf g} as ``markers".
While we stick to the current molecule either $\upr{a}{\lambda}$ or $\upr{\lambda}{t}$,
the observer maps the result (a molecule without markers) to $b$.
As soon as we attach to the current molecule a marker, the observer maps the resulting molecule to $d$,
until the strand with a marker is extended or until the molecule is completed.

Suppose that, when the first marker is attached, the length of the strand with that marker
is $l_1$, the length of the other strand is $l_2$ (clearly, $l_1>l_2$), and then
the output produced so far is $b^{l_1+l_2-2}d$. To complete the molecule by extending
the strand without the marker, we need to attach $l_1-l_2$ symbols to it,
and in this case the observer outputs $d^{l_1-l_2-1}\lambda$.
Thus, the resulting string $x$ consists of $l_1+l_2-2$ $b$'s and $l_1-l_2$ $d$'s.
Since $l_2\geq 1$, the difference between the number of $b$'s and the number of $d$'s
is $l_1+l_2-2-(l_1-l_2)=2l_2-2\geq 0$. (Recall that in case we attach a symbol to
a string with the marker, the observer only outputs $\lambda$, so the inequality
$m=|x|_b\geq |x|_d=n$ remains valid, and all the combinations $(m,n)$, $m\geq n$ are possible).
Hence, $L(\gamma)=L_1$.
 
 \bigskip 
 
 To get computational completeness  we do not need ``complicated" sticker systems but
simple regular sticker systems and  an observer that is able to
discard any ``bad" evolution. 

\begin{theorem}\label{univers} 
For each $L\in RE$ there exists an observable sticker system \label{universality}
$\phi$ such that $L_\bot(\phi)=L.$ \cite{AC04}
 \end{theorem}

Notice that, as already remarked earlier, using Theorem \ref{univers}, the definition of $L_\bot(\phi)$
and the fact that recursive languages are closed under
intersection with regular languages, we obtain:
\begin{corollary} \label{NotREC} 
There exists an observable sticker system $\phi$ such that $L(\phi)$ is a
non-recursive language. \cite{AC04}
\end{corollary} 

An interesting question here is why we obtain computational completeness with a regular system while with regular grammars the architecture is much weaker. Intuitively the reason for this is the fact that a grammar can only write a terminal once, while the sticker system can do so twice -- once in the upper strand, once in the lower. Thus the workspace is rewritable in a limited manner while for a regular grammar the workspace is essentially only its one non-terminal.

\section{Research Perspectives}

As the surveyed results show, the paradigm of evolution and observation is indeed quite universally applicable to discrete systems known in Theoretical Computer Science. As soon as a simple way of mapping its configurations to single letters is found, any such system can be the base of a computation by observation. However, not every type of system is equally adapt for achieving great computational power. Actually, Theorem \ref{th-ultra} shows that even a decrease in power can occur. Here the crucial factor seems to be space that can repetitively be rewritten; if it is finite, only finite-state computations seem possible.

Besides investigating further specific systems with respect to their power in an evolution-observation-system, it seems very interesting to try and formalize the thoughts of the preceding paragraph and to try to identify other such crucial factors. This could eventually lead to a type of measure for the capacity of systems to process information. On the other hand, limiting these crucial resources would lead to complexity hierarchies. A first result in this direction is that a linear space bound on a grammar's sentential form in an otherwise computationally complete model characterizes the context-sensitive languages \cite{CL05}.

In this paper we have presented only generative devices. However,  the paradigm can be applied also to construct accepting devices.  In this case, one can imagine introducing an input to the basic system, and observe its evolution. If the evolution of the observed system is the one expected (for example follows a certain pattern) then the input is accepted by the system, otherwise is considered to be rejected. This approach has been applied to string rewriting (\cite{SR}, where a characterization of context-sensitive languages is provided) and DNA splicing (\cite{SS}).

 Another possibility is to use the paradigm to formalize of the notion of ``abstraction''. In fact, the process of observation collapses the state-space of the basic system into a smaller one, doing a sort of abstraction on the observed system (keeping some specific information, while ignoring others). The observed behaviors are then ``upper approximations'' of the behaviors of the basic system. Two interesting problems should be investigated: (i) how to guarantee that the observed behaviors maintains some specific properties of the basic system (e.g., if the basic system oscillates then this is maintained in the observed behaviors); (ii) given the basic system and the observer, how to provide, in an algorithmically efficient manner, a finite description (e.g.,  a grammar) of the sets of observed behaviors.

One more application of the paradigm could  be in the area of quantum computation. There the concept of observer is crucial, and actually, a major problem in quantum experiments is to design the correct observer. However no notion of complexity of the external observer has been formally defined. In the proposed paradigm one could formalize this notion by explicitly dividing the observed system from the observer (\cite{caludePC}).

\bibliographystyle{eptcs}

\end{document}